\documentclass[12pt]{article}
\usepackage{graphicx}
\usepackage{epstopdf}
\usepackage{amsmath}
\usepackage{amsfonts}
\usepackage{amssymb}
\usepackage{color}
\usepackage{mathrsfs}

\newcommand{\be}{\begin{equation}}
\newcommand{\ee}{\end{equation}}
\newcommand{\bea}{\begin{eqnarray}}
\newcommand{\eea}{\end{eqnarray}}
\newcommand{\barr}{\begin{array}}
\newcommand{\earr}{\end{array}}

\def\beq{\begin{equation}}
\def\eeq{\end{equation}}
\def\be{\begin{equation}}
\def\ee{\end{equation}}
\def\bea{\begin{eqnarray}}
\def\eea{\end{eqnarray}}
\def\d{{\partial}}

\def\mpl{M_{\rm Pl}}

\begin{document}


\setcounter{page}{1} \baselineskip=15.5pt \thispagestyle{empty}

\begin{flushright}
\end{flushright}

\begin{center}

\def\thefootnote{\fnsymbol{footnote}}

{\Large \bf A Note on the Consistency Condition\\[0.3cm] of Primordial Fluctuations}
\\[0.5cm]

{\large Leonardo Senatore$^{1,2}$, and Matias Zaldarriaga$^3$}
\\[0.5cm]

{\normalsize { \sl $^{1}$ Stanford Institute for Theoretical Physics, Stanford University, Stanford, CA 94306}}\\
\vspace{.3cm}

{\normalsize { \sl $^{2}$ Kavli Institute for Particle Astrophysics and Cosmology, Stanford University and SLAC,\\ Menlo Park, CA 94025}}\\
\vspace{.3cm}

{\small \normalsize{\sl $^{3}$ Institute for Advanced Study, Einstein Drive,  Princeton, NJ 08540}}

\end{center}

\vspace{.8cm}

\hrule \vspace{0.3cm}
{\small  \noindent \textbf{Abstract} \\[0.3cm]
\noindent We show that the squeezed limit of $(N+1)$-point functions of primordial correlation functions in which one of the modes has a very small wavenumber can be inferred from the spatial variation of locally measured $N$-point function.  We then show how in single clock inflation a long wavelength perturbation can be re-absorbed in the background cosmology and how in computing correlation functions the integrals of the interaction Hamiltonian are dominated by conformal times of order of the short wavelength modes, when the long mode is already outside of the horizon. This allows us to generalize the consistency condition for $N$-point functions to the case in which the short wavelength fluctuations are inside the horizon and derivatives acts on them. We further discuss the consistency condition in the soft internal squeezed limit in which in an $(N+M)$-point function with $(N+M)$ short modes  the sum of the first $N$ modes is a very soft momentum. 
These results are very useful to study infrared effects in Inflation.}
 \vspace{0.3cm}
\hrule



\section{Introduction}

The consistency condition of the three-point function for single clock inflation was originally proposed in~\cite{Maldacena:2002vr,Creminelli:2004yq,Cheung:2007sv}. It showed that the three-point function in the kinematical limit in which one of the modes has wavelength much shorter than the others, the so-called squeezed limit, can be expressed in terms of the two-point function.

What is its origin and usefulness? The fact that in single clock inflation a consistency condition in the squeezed limit exists is due to two reasons. The first is that in single clock inflation the solution is an attractor. This implies that long wavelength fluctuations, as they are stretched to longer and longer wavelengths, become locally irrelevant, unobservable. This means that locally the dynamical behavior of the short distance fluctuations is not substantially affected by the long wavelength fluctuations. There are projection effects because  the coordinates in which in the universe looks unperturbed are different form the global coordinates used to describe the entire space-time. The consistency condition states nothing but this, and since we use comoving coordinates that are defined globally to describe the inflationary perturbations, such a trivial statement becomes non-trivial once expressed in terms of comoving coordinates. 

The second reason why the consistency condition holds is because a long wavelength fluctuation, at leading order in its small wavenumber, is not affected by short wavelength fluctuations. A long wavelength fluctuation is therefore completely frozen by the time shorter scale fluctuations come out of the horizon. If the long mode were to depend on the actual realization of the short modes, then the consistency condition would not hold. Such corrections are highly suppressed in the squeezed limit.

All of this means that a long wavelength fluctuation is locally unobservable, and this allows us to infer the $N$-point function in some squeezed limit from the $(N-1)$-point function. The properties that justify the existence of the consistency condition depend solely on the fact that locally a long wavelength mode, longer than the Hubble scale, is not observable. The short modes can be well inside the horizon. 
\begin{itemize}

\item One of the purposes of this paper is to prove that the consistency condition holds in the case the short wavelength fluctuations are still inside the horizon. 

\item Since now the short mode can be inside the horizon, we can allow for derivatives acting on them. We will therefore generalize the consistency condition to include derivatives acting on the short wavelength fluctuations. 

\item Furthermore, we will discuss the consistency condition in the case of correlation functions of arbitrary order and with an arbitrary number of external long wavelength fluctuations. We will also discuss a consistency condition that is valid for correlation functions where the sum over a subset of momenta is taken to be very small. We refer to this regime as `internal soft momenta', to distinguish it from the `external short momenta' we just discussed.

\item We state our results concentrating on fluctuations of scalar curvature $\zeta$ modes. They can be trivially extended to include either short or long wavelength tensor fluctuations.

\end{itemize}

We show the somewhat intuitive fact that it is possible to infer the squeezed limit of an $(N+1)$-point function with one external soft momenta $k_L$, from the  spatial variation  of locally-measured $N$-point functions on scales $k_L$:  
\be\label{squeeze_intro}
\langle {\hat P}_N(k_1,\cdots, k_N,k_L) \zeta(q_L) \rangle \approx (2 \pi)^3 {\delta^D}(k_L + q_L) Q_{N+1}(k_1,\cdots,k_N,k_L)\ ,
\ee
where $ {\hat P}_N(k_1,\cdots, k_N,k_L)$ represents the spatial variation on scales $k_L$ of the locally measured $N$-point function.
 Analogously,  we consider $(N+M)$-point functions in the so-called internal-squeezed-limit where all momenta are short-scale and locally-measurable,  but where the sum of the first $N$ momenta is equal to a very long wavelength mode $k_L$. We show that the $(N+M)$-point function in this kinematical limit can be inferred by correlating the spatial variations on the scale $k_L$ of locally-measured $N$ and $M$ point functions constructed with short-wavelength modes:
 \be\label{squeeze2_intro}
\langle {\hat P}_N(k_1,\cdots, k_N,k_L) {\hat P}_M(q_1,\cdots, q_M,q_L) \rangle \approx (2 \pi)^3 {\delta^D}(k_L + q_L) Q_{N+M}(k_1,\cdots,k_N,q_1,\cdots, q_M)\ .
\ee

We finish this introduction and summary by commenting on the usefulness of the consistency condition. 
\begin{itemize}
\item The initial motivation of checking the correctness of these difficult calculations will always  remain valid~\cite{Maldacena:2002vr}.

\item It offers a general characterization of the squeezed limit of $N$-point functions dependent only on lower order $N$-point functions. In general, it shows that the $N$-point function in the squeezed limit is connected to the scale dependence of lower order $N$-point function. This means that we expect a small effect on those scales for single clock inflation, all coming from projection effects, nothing from dynamical effects.

\item It offers a concise way to rule out single clock inflation: if we measure in the squeezed limit an $N$-point function that does not respect the consistency condition, it cannot be originated from single clock inflation. In reality, thanks to the Effective Filed Theory of Inflation~\cite{Cheung:2007st}~\footnote{See~\cite{Cheung:2007st,Cheung:2007sv,Senatore:2009cf,eft,loops1} for a sample of references related to this field.}, we can make much more powerful statements. A detection of {\it any} $N$-point function that is not reproducible by the Effective Field Theory of Inflation will rule out single clock inflation even if it satisfied the consistency condition in the squeezed limit. Even though this statement is much more powerful than the one coming from the consistency condition, still the squeezed limit is a kinematical regime that is very simple and connects very nicely to the bias and its scale dependence of collapsed objects~\cite{Dalal:2007cu}. 

\item At least to us, one of the most remarkable applications of the consistency condition is in exploring infrared effects in quantum loop computations in inflation. As it has been shown in~\cite{loops1} (see also~\cite{Senatore:2009cf}), when calculating loop corrections to inflationary observables, there are convolution integrals that are technically very challenging to compute. The consistency condition implies that the integrand resulting from the sum of many diagrams basically becomes in the squeezed limit a total derivative, allowing for a simpler calculation of the loops integral in this kinematical regime. This has allowed us to prove that there is no time-dependence in $\zeta$ when it is outside of the horizon even at loop order~\cite{loops1}. This is a very non-trivial statement, as naively many single diagrams induce a time dependence in $\zeta$~\cite{Weinberg:2005vy,Weinberg:2006ac,Kahya:2010xh}. It is only in the sum of many of them that the time dependence cancels out.

\end{itemize}

\section{The squeezed limit of $N$-point function}\label{squeeze-spatial-var}

Here we describe the relation between squeezed limit and spatial variation of lower order $N$- point functions.
This also includes cases when a combination of momenta tends to zero rather than one momenta. 

We start by dividing space into volumes centered around $x_\alpha$. For simplicity we will take all of these volumes to have the same shape and size but of course this is not necessary, we only do it to keep the notation simple. 

We are interested in the correlation functions as measured inside each of the volumes. First we will define the restricted Fourier transform,
\bea
\zeta_\alpha(k) &=& \int dx\ W(x-x_\alpha) \zeta(x)  e^{i k (x-x_\alpha)} \nonumber \\
&=& \int \tilde {dk^\prime}\  W(k^\prime)  \zeta(k-k^\prime) e^{-i (k-k^\prime)x_\alpha},
\eea
where $W$ is one inside the volume and zero outside, $\tilde {dk^\prime}= d^3{k}^\prime/(2 \pi)^3$ and we have used the same symbol for a function and its Fourier transform. 

The connected $N$-point functions of $\zeta$, $Q_N$, are defined as:
\be
\langle \zeta(k_1) \cdots  \zeta(k_N) \rangle_c = (2 \pi)^3 \delta^D(\sum_j k_j) \ Q_N(k_1,\cdots, k_N).
\ee
In each volume the measured $N$-point function is given by: 
\be
\langle \zeta_\alpha(k_1) \cdots  \zeta_\alpha(k_N) \rangle_c = (2 \pi)^3 \tilde{\delta}^D(\sum_j k_j) \ Q_N(k_1,\cdots, k_N),
\ee
where
\be
\tilde{\delta}^D(k)=\int dx\ W(x) e^{i k x}.
\ee
The function $\tilde{\delta}^D$ approximates a $\delta$-function but has a width that scales as $1/V$ and  $\tilde{\delta}^D(0)=V$

We will call $P_N(k_1,\cdots, k_N,x_\alpha)$ the estimate of $Q_N(k_1,\cdots, k_N)$ that one can obtain in the volume centered around $x_\alpha$,
\be
P_N(k_1,\cdots, k_N,x_\alpha) = {1\over V} \zeta_\alpha(k_1) \cdots  \zeta_\alpha(k_N).
\ee
We will also introduce its Fourier transform which can be computed by integrating the individual $P_N$ over the entire space:
\be
{\hat P}_N(k_1,\cdots, k_N,k_L)= \sum_\alpha V \  P_N(k_1,\cdots, k_N,x_\alpha) e^{i k_L x_\alpha}.
\ee

By simple manipulations of the above expressions one can show that
\be\label{squeeze1}
\langle {\hat P}_N(k_1,\cdots, k_N,k_L) \zeta(q_L) \rangle \approx (2 \pi)^3 {\delta^D}(k_L + q_L) Q_{N+1}(k_1,\cdots,k_N,k_L),
\ee
where we have assumed that $k_1, \cdots k_N$ are short modes and $k_L$ is long. By that we mean that $k_i^3 V \gg 1$ while $k_L^3 V \ll 1$.
In other words the $k_i$ modes are modes that can be well measured inside a volume of size $V$ while $k_L$ is almost constant over that volume. 
As a consequence $k_L \ll k_i$. 

Equation (\ref{squeeze1}) states that the squeezed limit of $Q_{N+1}$, where one of the modes is much longer than the others has a simple interpretation. It can be measured by studying how the estimates of the $N$-point function measured in small regions correlate with a long wavelength mode. This result will be useful for us because it implies that in order to make a theoretical prediction for the squeezed limit of $Q_{N+1}$ one only needs to be able to calculate $Q_N$ over small regions, regions over which the long modes is spatially constant.

There is another squeezed limit we can address. Assume we are interested in a situation in which a subset of the momenta almost add up to zero. For example in the case of a four point function this would be a situation in which all the sides of the quadrilateral are large but a diagonal is very small. We will label this by 
$Q_{N+M}(k_1,\cdots,k_N,q_1, \cdots q_M)$ where $\sum k_i = -\sum q_i \ll k_i,\ q_i$. 

Again simple manipulations of the above formulas show that $Q_{N+M}$ is related to the cross correlation between $P_N$ and $P_M$,
\be\label{squeeze2}
\langle {\hat P}_N(k_1,\cdots, k_N,k_L) {\hat P}_M(q_1,\cdots, q_M,q_L) \rangle \approx (2 \pi)^3 {\delta^D}(k_L + q_L) Q_{N+M}(k_1,\cdots,k_N,q_1,\cdots, q_M),
\ee
where again the squeezed limit has been assumed. 

In summary the squeezed limit of an $N$-point function both when one momenta is much smaller than the rest or when a subset of momenta sum to a very small value can be equivalently thought of as measure of the spatial variation of lower order $N$-point functions measured on smaller regions.

\section{Long mode in a different gauge}\label{gauge}

If we consider a set of $\zeta$ fluctuations that involve both long and short fluctuations, we aim here to provide a change of coordinates valid in a local patch that takes us from the metric written in standard $\zeta$ gauge to a form that is locally of the form of a homogeneous anisotropic universe. This shows that locally a long wavelength $\zeta$ fluctuations looks like an homogeneous anisotropic universe.  We start from the metric in ADM parametrization
\be
ds^2=-N^2 dt^2+\sum_{ij} \hat h_{ij}\left(dx^i+N^i dt\right)\left(dx^j+N^j dt\right)\ ,
\ee
where in this section we suspend the convention of summing over repeated indices.  
In $\zeta$ gauge the spatial metric takes the form $\hat h_{ij}=\delta_{ij}a(t)^2 e^{2\zeta}$.
We are going to perform the following change of coordinates 
\be
x^i=e^{\beta_{ij}(t)}\tilde x^j+C^i(t)\ .
\ee
that keeps the fluctuations of the inflaton zero, as it can be straightforwardly verified.
In this paper, we concentrate on tree-level correlation functions where we take all the momenta to be different. Because of this, each long-wavelength $\zeta$ mode enters linearly in the correlation functions, and  we can therefore work at linear order in the long modes. Further, we can use rotational invariance to consider a long mode with wavenumber only along the $\hat z$ direction,
\be
\zeta_L(\vec x,t)={\rm Re}\left[\zeta_0(t) e^{i k_L z}\right]\ .
\ee
We aim at bringing the metric to the following form
\bea\label{eq:anisotropic_metric}
&& \hat h_{11}=\hat h_{22}=e^{2\rho(t)+2\zeta_0(t)+2\lambda_0(t)} e^{2\zeta(\vec x,t)}\ , \\ \nonumber
&& \hat h_{33}=e^{2\rho(t)+2\zeta_0(t)-4\lambda_0(t)} e^{2\zeta(\vec x,t)}\ , \\ \nonumber
&& N_i=\d_i\psi(\vec x,t)\ , \\ \nonumber
&& N=1+\delta N_L(t)+\delta N(\vec x,t)\ ,
\eea
It will be enough to take $\beta_{ij}=\beta(t) \delta_{i3}\delta_{j3}$. 
The only subtle point in the change of variables that we are going to perform has to do with the constraint variables $N,N^i$. Their specific solution depends on the particular single field model considered, but the general features remain unchanged. For example, in the case of standard slow-roll inflation, at linear order in the long modes, we have 
\be
\vec N_L=\left\{0,0,{\rm Re}\left[i\frac{\dot H}{H^2}\frac{1}{k_L}\dot\zeta_0 e^{ik_L e^\beta\tilde z}\right]\right\}+{\cal O}(k_L\, \zeta_0)\ ,
\ee
which does not have a nice behavior for $k_L\rightarrow 0$. We need therefore to enforce that our change of coordinates not only fixes to zero $N^i$ at one point, say the origin, $N^i_0=0$, but also it must set to zero $\d_i N^j$ at the origin, $(\d_i N^j)_0=0$. This will guarantee that neglected terms are suppressed in the limit $k_L\rightarrow 0$. 

Simple algebra shows that the solution is
\bea
&&\vec C=\left\{0,0,-\int^t_{t_0} dt'\;\vec N_{L,0}(t')\right\}\  , \\
&&\beta=-\int^t_{t_0} dt'\; (\d_i N^i_L)_0(t')\ .
\eea
The metric then takes the form of (\ref{eq:anisotropic_metric}), with, in the new coordinates
\bea
\tilde N^i_{L,0}=0\ , && \qquad \left(\d_j \tilde N^i_L\right)_0=0\ , \qquad \\ \nonumber
  \tilde\zeta(\vec{\tilde x},t)=\zeta\left(\vec x(\vec{\tilde x},t),t\right)+\frac{2}{3} \int^tdt\;\frac{\dot H}{ H^2}\dot \zeta_0(\vec {\tilde x},t)\ ,&&\qquad\lambda_0(t)=-\frac{1}{3} \int^tdt\;\frac{\dot H}{ H^2}\dot \zeta_0\ ,
\eea
as we wanted to show.
$\delta\tilde N_L$ does not vanish in general, but it is proportional to $\dot\zeta_L$ as it can be readily inferred from the structure of the constraint equations. For standard slow roll inflation, at linear order, we have
\be
\delta\tilde N_L=\frac{3 H^2}{3 H^2+\dot H}\dot\zeta_L\ .
\ee
Notice that the short mode fluctuations $\zeta_S$ transform as a scalar under this change of coordinates
\be
\tilde\zeta_S(\vec{\tilde x},t)=\zeta_S\left(\vec{ x}(\vec{\tilde x},t),t\right)\ .
\ee

Let us see what the leading correction in $k_L$ is. If we substitute the linear solution for $\zeta$, as we are allowed to do when we consider the consistency condition, then we have that $\dot\zeta\propto k_L^2/a^2$. This means that all the time derivatives are of order $k_L^2$.  We can also perform a residual time-independet coordinate transformation to remove a constant gradient at the origin. Since at late times the time derivative of $\zeta$ goes as $k_L^2$, we can neglect the time dependence of the gradient of $\zeta$ up to cubic order in $k_L^2$. The additional change of coordinates we need to do is
\be
x^i=\tilde x^i- \vec\nabla_L\zeta|_0\cdot \vec{\tilde x} \ {\tilde x}^i+ {1\over 2} \vec\nabla_L\zeta|_0 \ {\tilde x}^2 \ \ .
\ee
At this point $\zeta$ has a Taylor expansion around the origin that starts with a constant and then has a term in $k_L^2$. We conclude that corrections to the homogenous metric are explicitly of order $k_L^2$. There is no correction linear in $k_L$.


The same procedure can be clearly performed at non-linear level in $\zeta_L$ using a generic matrix $\beta_{ij}$, but this is not necessary when all the long wavelength modes have different $k_L$.

\section{Consistency condition at tree level}

We will prove several consistency conditions valid at tree level for different type of $N$-point functions. In general we will be interested in $N$-point functions of the form
\bea\label{expect1}
&&\langle \Omega | {\cal D}_1\zeta_1(t) \cdots  {\cal D}_N \zeta_N(t) | \Omega \rangle =\\
\nonumber && \qquad\qquad \langle \Omega | {\cal D}_1\left[U_{I}^\dagger(t,-\infty) \zeta_{I1}(t)U_{I}^\dagger(t,-\infty) \right] \cdots {\cal D}_N\left[ U_{I}^\dagger(t,-\infty) \zeta_{IN}(t) U_{I}(t,-\infty)\right] | \Omega \rangle\ ,
\eea
where the subscript $_i$ represent the coordinate of evaluation of $\zeta$ and ${\cal D}$ represents a generic derivative operator, including the identity.  $| \Omega \rangle$ represents the vacuum of the interacting theory. On the right hand side we wrote the expression in terms of the fields in the interaction picture, $\zeta_{Ii}$.  and the evolution operator,
\bea
U_{I}(t,-\infty)&=&T [e^{-i \int_{-\infty}^t  dt_1 H_{I}(t_1)}] \nonumber \\
&=& 1 + (- i)^1 \int_{-\infty}^t  dt_1 \ H_{I}(t_1) + (- i)^2 \int_{-\infty}^t  dt_1 \int_{-\infty}^{t_1}  dt_2 \ H_{I}(t_1)H_{I}(t_2) \nonumber \\
&+& (- i)^3 \int_{-\infty}^t  dt_1 \int_{-\infty}^{t_1}  dt_2 \int_{-\infty}^{t_2}  dt_3 \ H_{I}(t_1)H_{I}(t_2) H_{I}(t_3) + \cdots,
\eea
where $T$ stands for $T$-ordering and $H_{I}$ is the interaction hamiltonian. We also have
\bea
U_{I}^\dagger(t,-\infty)&=&\bar T [e^{+i \int_{-\infty}^t  dt_1 H_{I}(t_1)}],
\eea
where $\bar T$ stands for anti-$T$-ordering.

Finally equation (\ref{expect1}) involves the expectation value in the interacting vacuum. We can accomplish this by rotating the time integration adding a small imaginary part to the time variable and computing the expectation value in the vacuum of the free theory $|0\rangle$, 
\be
\langle \Omega | U_{I}^\dagger(t,-\infty) \zeta_{I1}(t) \cdots \zeta_{IN}(t) U_{I}(t,-\infty) | \Omega \rangle=\langle 0 | U_{I}^\dagger(t,-\infty_+) \zeta_{I1}(t) \cdots \zeta_{IN}(t) U_{I}(t,-\infty_+) | 0 \rangle
\ee

In fact it will be more convenient for us to rotate the contour of integration to be completely in the complex plane and write the integration variable $t_1= t + i x$ with $x$ running from $-\infty$ to 0. Note that the contour of integration in $U_{I}^\dagger(t,-\infty_+)$ needs to be rotated in the opposite direction, $t_1= t - i x$. Although the result in independent of the amount of rotation, the rotation by ninety degrees will make our arguments more transparent. Both the operators $\zeta_{Ii}$ and $H_I$ are built of operators in the interaction picture which are written in terms of creation and annihilation operators and solution to the free wave equation. For example for $\zeta(k,t)$ we have
\be
\zeta(k,t) = \zeta^{cl}(k,t) a_k +  \zeta^{cl*}(k,t) a_{-k}^\dagger.
\ee
When the mode is inside the horizon $k/(a(t)H(t))\gg 1$, the $\zeta(k,t)^{cl}$ has a WKB form, and matches to the positive frequency solution in flat space. As a result of our rotation in the countour of integration the oscillations of $\zeta^{cl}(k,t)$ become a decaying exponential. The choice of countours for $U_{I}(t,-\infty_+)$ and $U_{I}^\dagger(t,-\infty_+)$ guarantees that this is true for wavefunctions resulting from the expansion of both of these operators. 

We will evaluate $\langle 0 | U_{I}^\dagger(t,-\infty_+) \zeta_{I1}(t) \cdots \zeta_{IN}(t) U_{I}(t,-\infty_+) | 0 \rangle$ perturbatively. Each term in the expansion can be represented using a diagram as shown for example in Fig.~\ref{fig:consistency_external}. Vertices in the diagram come from the expansion of the evolution operators. Vertices might come from  $U_{I}$ or $U_{I}^\dagger$. Lines connecting the vertices or the vertices with the external operators lead to $\zeta^{cl}(k,t_a) \zeta^{cl*}(k,t_b)$ where $t_{a,b}$ are the time of evaluation of the operators $\zeta_{Ii}(t)$ or of the Hamiltonian $H_I$. It is important to notice that since the rotation of the contour of integration is done in opposite directions for $U_{I}$ and $U_{I}^\dagger$, the wave functions of the operators originating from the expansion of $U_I$ and $U_I^\dagger$ all decay exponentially at early times.

We could compute the expectation value around a perturbed background characterized by a classical $\zeta_B$,
\be\label{expect-classical}
\langle \Omega | \zeta_1(t) \cdots \zeta_N(t) | \Omega \rangle_{\zeta_B},
\ee
the same perturbative expansion and diagrams apply in this case with the exception that there are lines connecting the vertices to the classical perturbation 
$\zeta_B$. In this case, $\zeta_B$ is a real function, the same function when inserted in vertices. 

\subsection{One soft external momentum}

The first case we consider is the expectation
\be
\langle \Omega | \zeta(k_1,t) \cdots \zeta(k_N,t) \zeta(k_L,t) | \Omega \rangle= (2 \pi)^3 \delta^D(\sum k_i + k_L) Q_{N+1}(k_1,\cdots, k_N, k_L),
\ee
where we will assume that $k_L$ is the softest momenta (longest wavelength) and it is outside the horizon at time $t$, $k_L/( a(t)H(t)) \ll 1$. By softest momenta we mean not only that $k_L \ll k_i$ but also all partial sums of the $k_i$ (except of course the sum of all of them which by momentum conservation is equal to $k_L$). 

Due to the rotation of the contour, any vertex connected to one of the $\zeta(k_i,t)$ cannot be separated in time from $t$ by more than a conformal time difference of $\delta \eta \sim 1/k_i$ before the corresponding wave function starts decreasing exponentially. Lines connecting two vertices cannot be too long either, if the momenta flowing in that line is $k$ then the conformal time difference between the vertices cannot be larger than $\delta \eta \sim 1/k$. In fact because $k_L$ is the smallest momenta even among the partial sums of the $k_i$, all the vertices in the diagram have to be close to the final time, with conformal time difference of order $1/k_S$ where $k_S$ generically represents the size of the short momenta~\footnote{This is the step that fails when considering correlation functions at loop level. In that case line connecting vertexes that are not directly connected to the external $\zeta(k_i)$, can have arbitrarily low momenta. This prevents us from extending the proof of the consistency condition at loop level, though the consistency condition is expected to hold at any order in perturbation theory.}. 

\begin{figure}[h]
\begin{center}
\includegraphics[width=17cm]{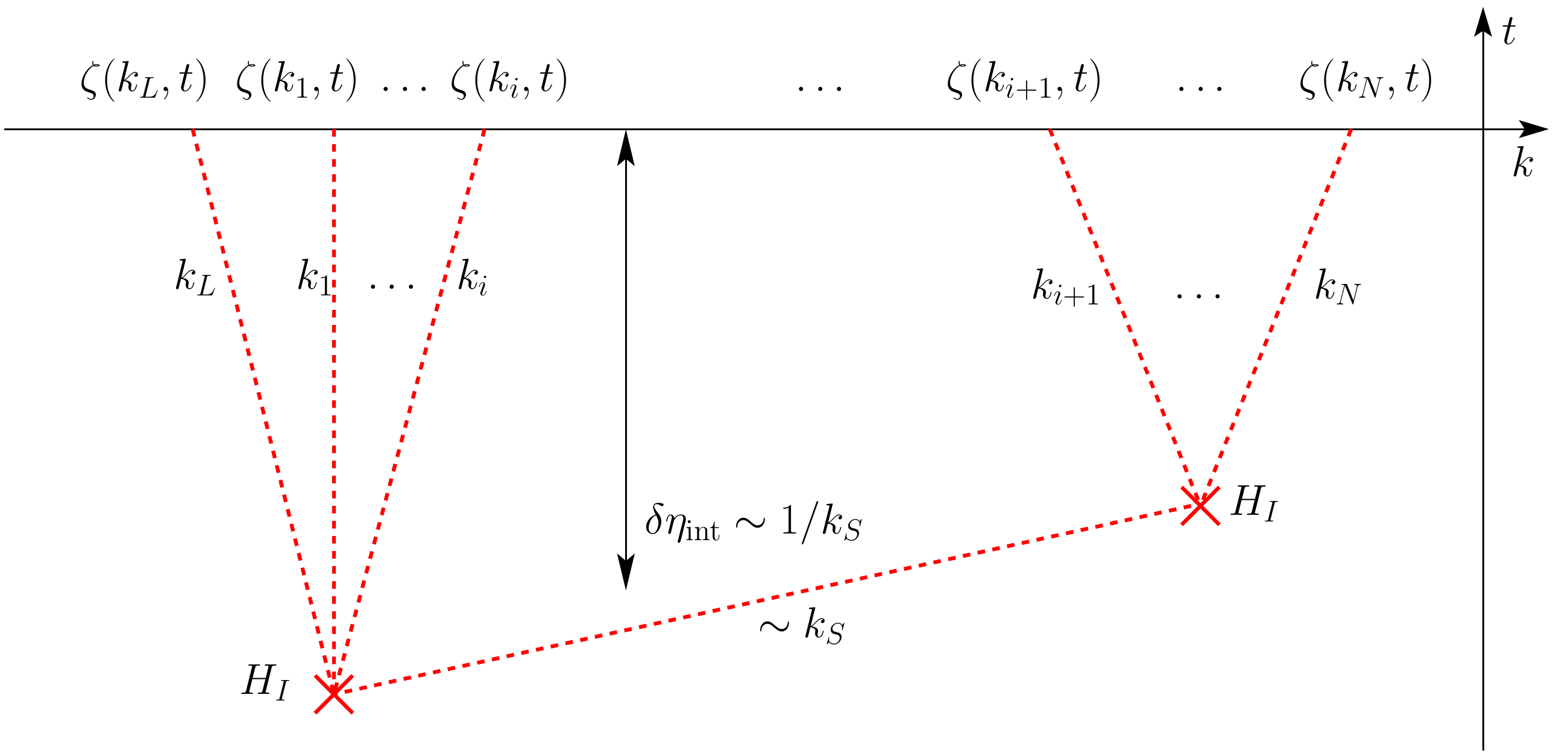}
\caption{\label{fig:consistency_external} \small\it Consistency condition with one soft external momentum. Dashed red lines connecting the vertices or the vertices with the external operators lead to $\zeta^{cl}(k,t_a) \zeta^{cl*}(k,t_b)$ where $t_{a,b}$ are the time of evaluation of the operators $\zeta_{ki}(t)$ or of the Hamiltonian $H_I$. Because of the contour rotation, the wavefunctions of operators with momentum $k$ associated to the interaction Hamiltonians decay at early time with a time scale of order $\delta \eta\sim 1/k$. Therefore, all the vertices must be close in time to the final time, within a $\delta\eta\sim 1/k_S$. This means that the long wavelength fluctuation $\zeta_{k_L}$ is out of the horizon both at the final time and at the time of evaluation of the integrals. This means that at all times involved in the calculation $\zeta_{k_L}$ can be treated as a rescaling of the coordinates, implying the consistency condition. This also means that the result would have not changed if long wavelength fluctuation $\zeta(k_L)$, which in the plot is represented as evaluated at the same final time as the short wavelength fluctuations, had been evaluated at a different final time when the mode is still outside of the horizon.}
\end{center}
\end{figure}

Our assumption that $k_L$ was already well outside the horizon at the final time and the fact that all vertices are within $1/k_S$ of the final time implies that $k_L$ is also well outside the horizon at the time when all the vertices are computed. The $k_L$ wave function is real and thus $k_L$ can be thought of a classical background.  This is represented in Fig.~\ref{fig:consistency_external}.

In section \ref{squeeze-spatial-var} we showed that the squeezed $N$-point function we are considering here can be also obtained by computing $\zeta(k_1,t) \cdots \zeta(k_N,t)$ in small volumes and then correlating its spatial variations with $\zeta(k_L,t)$. In section \ref{gauge} we showed that over such a small volume one can perform a gauge transformation that makes the long mode look like a homogenous anisotropic universe. Once the mode is outside the horizon though, the anisotropic expansion has decayed and the only residual effect of the long mode is equivalent to a rescaling  $x^i\rightarrow x^i \exp[\zeta(k_L,t)]$.

Thus the diagrams we are computing can be obtained simply by computing the Fourier transform of the expectation value
\be
\langle \Omega | \zeta(x_1,t) \cdots \zeta(x_N,t) | \Omega \rangle_{\zeta_L}\ ,
\ee
the $N$-point function in the background of $\zeta(k_L,t)$ which is simply given by the answer in the absence of $\zeta(k_L,t)$ but with rescaled momenta $k_i\rightarrow k_i \exp[-\zeta(k_L,t)]$ and multiplied by a factor of $\exp[-D_{Q_N}\cdot \zeta(k_L,t)]$, with $D_{Q_N}$ being the dimension in units of length of $Q_N$~\footnote{This means that two operators involving $\zeta$ or $\dot\zeta$ have the same spatial dimensions $D_{Q_N}$.}. This last factor simply follows from the integrals in $d^3x_i\,,\ i=1,\ldots, N$ associated with the Fourier transform, after taking into account that one integration gives the momentum conserving delta function. In formulas:
\bea
&&\!\!\!\!\!\!\!\!\!\!\langle \Omega | \zeta(k_1,t) \cdots \zeta(k_N,t) | \Omega \rangle_{\zeta_L}=\int d^3x_1\ldots d^3x_N \langle\zeta(x_1)\ldots\zeta(x_N)\rangle_{\zeta_L} e^{i k_1 x_1+\ldots i k_N x_N}=\\ \nonumber
&&\!\!\!\!\!\!\!\!\!\!(2\pi)^3 \delta^D(k_1+\ldots +k_N)\int d^3\Delta x_1\ldots d^3\Delta x_{N-1} \langle\zeta(\Delta x_1 e^{\zeta_L})\ldots\zeta(\Delta x_{N-1} e^{\zeta_L})\zeta(0)\rangle e^{i k_1 \Delta x_1+\ldots i k_{N-1} \Delta x_{N-1}}\\ \nonumber
&&\!\!\!\!\!\!\!\!\!\!=(2\pi)^3 \delta^D(k_1+\ldots +k_N) e^{-3(N-1)\zeta_L}Q_N(k_1 e^{-\zeta_L},\ldots, k_N e^{-\zeta_L})\\ \nonumber
&&\!\!\!\!\!\!\!\!\!\!=(2\pi)^3 \delta^D(k_1+\ldots +k_N) e^{-D_{Q_N}\zeta_L}Q_N(k_1 e^{-\zeta_L},\ldots, k_N e^{-\zeta_L})\ .
\eea
For an $N$-point involving only $\zeta$s and $\dot\zeta$s we  have  $D_{Q_N}=3(N-1)$. 
The correlation function we seek is obtained by expanding this function to linear order in $\zeta(k_L,t)$, multiplying the answer by $\zeta(k_L,t)$ and taking expectation value over the long mode. We get
\be
Q_{N+1}(k_1,\cdots, k_N, k_L) \approx- {\tilde\partial Q_{N}(k_1,\cdots, k_N) \over \tilde\partial \ln k} P(k_L)\ ,
\ee
where $P$ is the power spectrum of $\zeta$ and
\be\label{eq:dtilde}
{\tilde\partial Q_{N}(k_1,\cdots, k_N) \over \tilde\partial \ln k}={\partial Q_{N}(k_1,\cdots, k_N) \over \partial \ln k}+D_{Q_N} Q_{N}(k_1,\cdots, k_N)\ .
\ee
Note that at no point have we assumed that the $k_i$ are outside the horizon. The above arguments are valid even if we are computing a correlation function during inflation and some or all of the short modes are inside the horizon.


\subsection{One soft internal momentum}

\begin{figure}[h]
\begin{center}
\includegraphics[width=17cm]{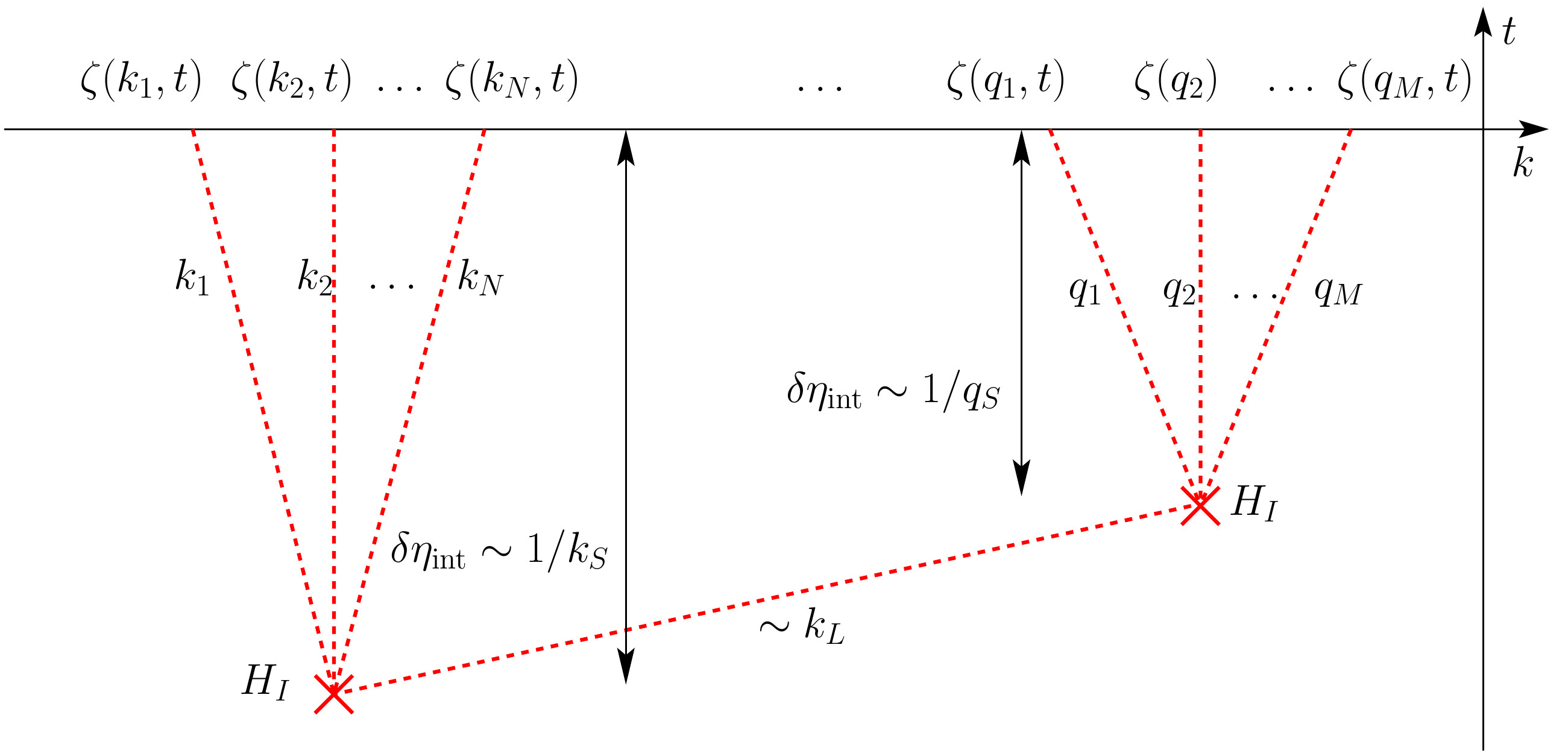}
\caption{\label{fig:consistency_middle} \small\it Consistency condition with one soft internal momentum: in this case, for the particular momenta configurations that we have chosen, there are some diagrams where one of the lines connecting two vertices has a very low momentum. These are the diagrams that dominate the $N$-point functions. Since all the vertices are connected to an external $\zeta$ which has a short momentum, the interaction has to happen at a time close to the final time by a time $\delta\eta\sim 1/k_S$, with $k_S$ being the typical momentum of the short fluctuations. This means that at the time of evaluation of the integrals the long mode is already outside of the horizon and can therefore be re-absorbed with a rescaling of the coordinates. This implies that the consistency condition holds.}
\end{center}
\end{figure}

We now want to consider 
\bea
\langle \Omega | \zeta(k_1,t) \cdots \zeta(k_N,t) \zeta(q_1,t) \cdots \zeta(q_M,t) | \Omega \rangle &=&(2 \pi)^3 \delta^D(\sum k_i +\sum q_j ) \nonumber \\ &&Q_{N+M}(k_1,\cdots, k_N,q_1,\cdots, q_M)\ ,
\eea
where the partial sums $\sum k_i = - \sum q_i\equiv k_L$ is the softest momenta but all of the $k_i$, $q_i$ and other partial sums are hard. We consider cases in which the soft internal momenta $k_L$ is outside the horizon. The dominant terms in this computation comes from diagrams such as the ones illustrated in Fig.~\ref{fig:consistency_middle}, in which all the $k_i$ momenta are connected among themselves, all the $q_i$s are connected among themselves and the two subdiagrams are connected by a single line. Only in that case is the momentum running in a line soft so this leads to the biggest contribution. This effect was first noticed in~\cite{Seery:2008ax} for the case of the four-point function for the graviton exchange, and was later generalized to arbitrary correlation functions with an internal momentum in~\cite{Leblond:2010yq}. Here we point out how it holds in the case the short wavelength fluctuations are still inside the horizon, derivative operators act on them and there are many soft internal momenta. 

All the vertices in the $k$ subdiagram are required to be within $\delta\eta \sim 1/k$ of the $t$ while all the vertices in the $q$ subdiagram are required to be within $\delta\eta \sim 1/q$ of $t$. Because $k_L$ is well outside the horizon at $t$ and all the vertices are so close to $t$, again the $k_L$ wave function is real and thus $k_L$ can be thought of a classical background for both the $k$ and $q$ subdiagrams.

In section \ref{squeeze-spatial-var} we showed that the squeezed $N$-point function we are considering here can be also obtained by computing $\zeta(k_1,t) \cdots \zeta(k_N,t)$ is small volumes and then correlating its spatial variations with $\zeta(q_1,t) \cdots \zeta(q_M,t)$. In section \ref{gauge} we showed that over such a small volume one can perform a gauge transformation that makes the long mode look like a homogenous anisotropic universe. Once the mode is outside the horizon though, the anisotropic expansion has decayed and the only residual effect of the long mode is equivalent to a rescaling $a(t)\rightarrow a(t) \exp[\zeta(k_L,t)]$.

Thus the diagrams we are computing can be obtained simply by computing the expectation values
\be
\langle \Omega | \zeta(k_1,t) \cdots \zeta(k_N,t) | \Omega \rangle_{\zeta_L} \ \ \ \& \ \ \  \langle \Omega | \zeta(q_1,t) \cdots \zeta(q_M,t) | \Omega \rangle_{\zeta_L},
\ee
the $N$ and $M$-point function in the background of $\zeta(k_L,t)$ which are simply given by the answer in the absence of $\zeta(k_L,t)$ but with rescaled momenta $k_i\rightarrow k_i \exp[-\zeta(k_L,t)]$ and $q_i\rightarrow q_i \exp[-\zeta(k_L,t)]$. The correlation function we seek is obtained by expanding these functions to linear order in $\zeta(k_L,t)$ and taking expectation value over the long modes. We get
\be
Q_{N+M}(k_1,\cdots, k_N, q_1,\cdots, q_M) \approx {\tilde\partial Q_{N}(k_1,\cdots, k_N) \over \tilde\partial \ln k} {\tilde\partial Q_{M}(q_1,\cdots, q_M) \over \tilde\partial \ln q} P(k_L)\ .
\ee

Note that at no point have we assumed that the $k_i$ or $q_i$ are outside the horizon. The above arguments are valid even if we are computing a correlation function during inflation and some or all of the short modes are inside the horizon.

\subsection{Many soft momenta}

Let us start consider the expectation with many soft external momenta
\bea
\langle \Omega | \zeta(q_1,t) \cdots \zeta(q_N,t) \zeta(k_{L1},t) \cdots \zeta(k_{LM},t)  | \Omega \rangle &=& (2 \pi)^3 \delta^D(\sum q_i + k_{Lj}) 
\nonumber \\ && Q_{N+M}(q_1,\cdots, q_N,k_{L1},\cdots, k_{LM}),
\eea
where we will assume that $k_{Li}$ are all soft momenta (longest wavelength) and outside the horizon at time $t$, $k_{Li}/( a(t)H(t)) \ll 1$. By softest momenta we mean not only that $k_{Li} \ll q_i$ but also of all partial sums of the $q_i$.  

As before the dominant terms in this expansion will come when there is a sub-diagram that connects all of the $q_i$ together which then attached by soft lines to the $k_{Li}$. Again all the vertices in the $q$ subdiagram will be close in time to the final time, with a maximum conformal time difference of order $\delta\eta\sim 1/q$. By assumption then all the soft modes attached to these vertices are well outside the horizon and can be thought as a classical rescaling of the coordinates in the $q$ subdiagram. Restricting to the connected component, we thus have a formula of the form
\be
Q_{N+M}(q_1,\cdots, q_N, k_{L1},\cdots, k_{LM}) \approx-{\tilde\partial Q_{N}(q_1,\cdots, q_N) \over\tilde \partial \ln k}Q_{M+1}(k_{L_1},\ldots,k_{L_M},-(k_{L_1}+\ldots+k_{L_N}))\ ,
\ee
Let us give an example. If we consider a $N$-point function with two additional soft lines we would write:
\bea
&&Q_{N+2}(q_1,\cdots, q_N, k_{L1},k_{L2}) \approx- {\tilde\partial Q_{N}(q_1,\cdots, q_N) \over \tilde\partial \ln k} Q_3(k_{L1},k_{L2},-(k_{L1}+k_{L2}))
\eea
In the case in which all modes are outside of the horizon, this generaliztaion was discussed first in~\cite{Huang:2006eha}.
In the case in which there are many soft internal momenta, specified by the partial sums
\be
\sum_{i=1}^{M_j} k_{p_{N,j}(i)}=-\sum_{i={M_j+1}}^{N} k_{p_{N,j}(i)}=k_{Lj}
\ee
with $p_{N,j}$ being a permutation of the $N$ momenta whose first $M_j$ element have a low partial sum denoted by $k_{Lj}$, the consistency condition becomes simply
\be
Q_{N}(k_1,\cdots, k_{N}) \approx \sum_{j}{\tilde\partial Q_{M_j}(k_{p_{N,j}(1)},\cdots, k_{p_{N,j}(M_j)}) \over \tilde\partial \ln k} {\tilde\partial Q_{N-M_j}(k_{p_{N,j}(M_j+1},\cdots, k_{p_{N,j}(N)}) \over \tilde\partial \ln k} P(k_{Lj})\ .
\ee

\subsection{Consistency condition for derivative operators}

So far we have concentrated on the case where we compute correlation functions of $\zeta$ operators with no derivative acting on them. Let us briefly discuss what happens when they are included. 

In the case of  time derivatives, the consistency condition holds unchanged. The only subtlety concerns the actual computation. From (\ref{expect1}) we can see that when the time derivative acts on the time evolution operator, vertices become a contact term of the form $[H_{I}(t),\zeta(t)]$, without the usual associated time integration:
\bea\label{eq:dot}
\dot \zeta(t) &=& \partial_t\left(U_{int}^\dagger (t,-\infty_+)\zeta_I(t) U_{int}(t,-\infty_+)\right)\simeq\\ \nonumber
&& i [H_{int}(t), \zeta_I(t)] +U_{int}^\dagger (t,-\infty_+)\dot\zeta_I(t) U_{int} (t,-\infty_+)\ .
\eea
An example of verification of the consistency condition for a derivative operator is given in App.~A~of~\cite{loops1}.

Instead in the case the derivative is a spatial derivative, then it simply comes out of the expectation value and acts on the final result. As we saw, the consistency condition tells us that in the squeezed limit the effect of the evolution of the operators is simply to enforce a rescaling of the spatial coordinates: $\zeta(k_i)\rightarrow \zeta(k_i\, {\rm Exp}[-\zeta_L])$. Since spatial derivatives can be safely moved out of the expectation value, they are not rescaled.  Notice that this factor is automatically taken into account by our definition of $\tilde\partial/\tilde\d\log k$, as $\d k_i/\d \ln k=k_i$ and in this case the spatial dimension of $Q_N$ changes accordingly. This means that the consistency condition for the case of one soft external momentum still reads
\be
Q_{N+1}(k_1,\cdots, k_N, k_L) \approx - {\tilde\partial Q_{N}(k_1,\cdots, k_N) \over \tilde\partial \ln k}  P(k_L) \ .
\ee
For the case one soft internal momentum, we analogously have
\bea
&&\!\!\!\!\!\!\!\!\!\!Q_{N+M}(k_1,\cdots, k_N, q_1,\cdots, q_M) \approx {\tilde\partial Q_{N}(k_1,\cdots, k_N) \over\tilde \partial \ln k} {\tilde\partial Q_{M}(q_1,\cdots, q_M) \over \tilde\partial \ln q} P(k_L)
\eea
Similar relations hold for the case of many soft momenta.
Examples of verification of the consistency condition for short modes inside the horizon and including operators involving space derivatives are given in the App.~\ref{app:consistency-inside}.

\subsection{Corrections to the Leading Behavior}

The consistency conditions that we have written provide us the behavior of the connected correlation function in the extreme squeezed limit when the soft momenta $k_L\rightarrow 0$. It is interesting to check what are the leading corrections, which scale as $k_L/k_S$. 

For the three-point function it has been proven in~\cite{Ganc:2010ff,Creminelli:2011rh} that the calculation scales as $k_L^2/k_S^2$, with the linear correction canceling. Here we provide a very simple way to see how the result can be generalized to all correlation functions. 

In section \ref{squeeze-spatial-var} we showed that the squeezed $N$-point function can be also obtained by computing $\zeta(k_1,t) \cdots \zeta(k_N,t)$ in small volumes and then correlating its spatial variations with $\zeta(q_1,t) \cdots \zeta(q_M,t)$. This is true up to effects of order $1/V\sim k_L^3$. In sec.~\ref{gauge} we have shown that over such a small volume we can find a local gauge transformation where the long mode appears as a local homogenous universe with corrections that are explicitly of order $k_L^2$. This shows that the only corrections linear in $k_L$ can come only from the change of coordinates from the local to the global frame, which indeed has terms linear in $k_L$. This would allow us to extend the consistency condition to include terms that are subleading in one power of $k_L$. This approach has just been developed in~\cite{Creminelli:2012ed}, with a slightly different language, and therefore we do not develop it further.

\section{Consistency condition at loop level: IR effects}

Our study of the case with many soft external momenta allows us to infer the existence of large IR effects in $N$-point functions at loop level although these effects are not physical but in a sense an artifact of the choice of coordinates. 

Loop correction to the $N$-point function when very soft modes are running in the loop are intimately related to the case we have computed. Rather than assume the long modes have been measured one needs to average over the unobserved amplitude of the long modes. 
Basically in the loop calculation pairs of soft external momenta will be closed together rather than being attached to external lines and the result integrated over the momenta in the loop. 

Because of the absence of the external wavefunctions, the four wavefunction in the pair collapse to two, which is nothing but the power spectrum of the long modes. Then the integral over momenta will lead to a logarithmic IR divergence~\cite{Giddings:2010nc,Gerstenlauer:2011ti}. 

These divergences are clearly not there when distances are measured in physical rather than comoving coordinates~\cite{Senatore:2012nq}. Our entire calculation was relied on the fact that the $N$-point function simply has the comoving momenta rescaled in the presence of the background mode, rescaled in such a way that the $N$-point function in physical units is unperturbed. 

The consistency condition is also useful when very high momentum modes are running in the loops. In this case the short modes are at horizon crossing or inside the horizon, and the integrals are technically very challenging to compute. One basically has to compute the perturbation due to a long wavelength mode of the product of two short wavelength modes, and then integrate over the short wavelength momentum. The consistency condition implies that the integrand resulting from the sum of many diagrams basically becomes in the squeezed limit a total derivative, allowing for a simpler calculation of the loops integral in this kinematical regime. This has allowed us to prove that there is no time-dependence in $\zeta$ when it is outside of the horizon even at loop order~\cite{loops1}. This is a very non-trivial statement, as naively many single diagrams induce a time dependence in $\zeta$~\cite{Weinberg:2005vy,Weinberg:2006ac,Kahya:2010xh}. It is only in the sum of many of them that the time dependence cancels out.



\section{Conclusions}

In single field inflation $N$ point functions satisfy certain consistency conditions when some of the momenta are very soft. These conditions are basically a consequence of the attractor nature of the inflationary solution, they are the translation into comoving coordinates of the fact that modes outside the horizon become locally unobservable.  These consistency condition are still valid even at times  when some of the modes are inside the horizon or have derivatives acting on them. These more general forms of the consistency condition are relevant for computing loop correction to inflationary correlation functions. 

\subsubsection*{Acknowledgments}

 L.S.~is supported by the National Science Foundation under PHY-1068380.
M.Z. is supported by the National Science Foundation under PHY-
0855425 and AST-0907969 and by the David and Lucile 
Packard Foundation and the John D. and Catherine~T.~MacArthur~Foundation. 

\begin{appendix}
\section*{Appendix}
\section{Consistency Condition inside the Horizon\label{app:consistency-inside}}

In this Appendix we discuss the three-point function in the squeezed limit in which one of the modes is much longer than the other two, but the other two are possibly still inside the horizon. We will verify that the consistency condition also holds in this regime. We will do this at leading order in slow roll parameters working in the case of standard slow roll inflation for some specific operator. 

For the case in which the short modes are still inside the horizon, the proof at leading order in slow roll parameters is very easy. In fact, contrary to what happens when we are interested in computing the correlation function of modes at a time when they are outside the horizon, in this case the leading interaction is of zero$^{th}$ order in the slow roll parameters. Indeed, it is not true that the $\zeta$ cubic action starts at first order in slow roll parameters (relative to the quadratic action). This is so only up to terms that can be removed by a field redefinition and that can therefore be evaluated at the final time. For modes that are outside of the horizon at the time of evaluation, these vanish. For modes that are not yet outside of the horizon, they do not, and they therefore represent the leading contribution in the slow roll expansion.

Following~\cite{Maldacena:2002vr}, the term we are discussing comes from the field redefinition:
\be
\zeta=\zeta_n+\frac{\zeta\dot\zeta}{H}+\ldots\ ,
\ee
where $\ldots$ represent terms suppressed by slow roll parameters. The variable $\zeta_n$ has a cubic action that is suppressed by slow roll parameters, and so negligible. At this point computing the three-point function is very straightforward. In the limit in which the long mode $k_3$ is much longer than the horizon $k_3/a(\eta)\ll H$ and $k_3\ll k_2\simeq k_1$, we have
\bea
&&\langle\zeta_{k_1}(\eta)\zeta_{k_2}(\eta)\zeta_{k_3}(\eta)\rangle\simeq (2\pi)^3\delta^{(3)}(\vec k_1+\vec k_2+\vec k_3) \frac{1}{H}\langle\dot\zeta_{k_1}\zeta_{k_1}+\zeta_{k_1}\dot\zeta_{k_1}\rangle'\;\langle\zeta_{k_3}^2\rangle'\\ \nonumber &&
\qquad=(2\pi)^3\delta^{(3)}(\vec k_1+\vec k_2+\vec k_3)\frac{1}{H} \d_tP(k_1,t)\;P(k_3)\ , \qquad\quad\ k_1\ll k_3\ ,
\eea
where the $\langle\rangle'$ symbol stays for the fact that we have removed the delta function from the expectation value. Using the wavefunction of the modes at leading order in slow roll parameters
\be\label{eq:zetacl}
\zeta^{cl}_k(\eta)=\frac{H}{2\sqrt{ \epsilon}\mpl}\frac{1}{k^{3/2}}\left(1-i k\eta\right)e^{i k \eta}\ ,
\ee
where $\epsilon$ is the slow roll parameter $\epsilon=-\dot H/H^2$, we obtain
\be\label{eq:consistency_inside}
\langle\zeta_{k_1}(\eta)\zeta_{k_2}(\eta)\zeta_{k_3}(\eta)\rangle\simeq-\frac{ H^4}{8  \mpl^4 \epsilon^2}\cdot\frac{\eta ^2}{k_1 k_3^3}\ .
\ee

In order to satisfy the consistency condition, the above result should be equal to
\be
\langle\zeta_{k_1}(\eta)\zeta_{k_2}(\eta)\zeta_{k_3}(\eta)\rangle\simeq- (2\pi)^3\delta^{(3)}(\vec k_1+\vec k_2+\vec k_3)\frac{\tilde\d P(k_1,\eta)}{\tilde\d\log k_1}P(k_3)\ .
\ee
Notice that since the short modes are still inside the horizon, their power spectrum is not yet scale invariant, so $\tilde\d P(k_1)/\tilde\d\log k_1$ is not slow-roll suppressed. Upon substitution of (\ref{eq:zetacl}), this is indeed equal to (\ref{eq:consistency_inside}), verifying the consistency condition for modes inside the horizon.

\subsection{Consistency condition for  operators with spatial derivatives}
  
Let us now consider the three-point function in the same regime of momenta as above for a spatial derivative operator of the form
\be
\left\langle\frac{1}{a(\eta)^2}\d_i\zeta_{k_1}(\eta)\d_i\zeta_{k_2}(\eta)\zeta_{k_3}(\eta) \right\rangle  \ .
\ee
Since when we compute the three-point function we simply evolve the operators and not their spatial derivatives, the result can be trivially obtained from the one above in eq.~(\ref{eq:consistency_inside}) to be
\bea \nonumber
&&\left\langle\frac{1}{a(\eta)^2}\left(\d_i\zeta\right)_{k_1}(\eta)\left(\d_i\zeta\right)_{k_2}(\eta)\zeta_{k_3}(\eta)\right\rangle \simeq (2\pi)^3\delta^{(3)}(\vec k_1+\vec k_2+\vec k_3)\;\frac{k_1^2}{a(\eta)^2}\frac{1}{H} \d_tP(k_1,t)\;P(k_3)\\
&&\qquad=-\frac{ H^6}{8 \mpl^4 \epsilon ^2}\cdot\frac{\eta ^4\,k_1}{k_3^3}\ , \qquad\quad\ k_1\ll k_3\ .
\eea
Notice that this operator does not satisfy a naive consistency condition, that would read
\bea\label{eq:consistency_inside_derivative_wrong}
&&\!\!\!\!\!\!\!\!\left\langle\frac{1}{a(\eta)^2}\left(\d_i\zeta\right)_{k_1}(\eta)\left(\d_i\zeta\right)_{k_2}(\eta)\zeta_{k_3}(\eta)\right\rangle\simeq- (2\pi)^3\delta^{(3)}(\vec k_1+\vec k_2+\vec k_3)\\ \nonumber \\ \nonumber
&&\qquad\qquad\frac{1}{a(\eta)^2}\frac{\tilde\d\left[k_1^2P(k_1,\eta)\right]}{\tilde\d\log k_1}P(k_3)=-\frac{H^6 }{8   \mpl^4 \epsilon ^2}\frac{\eta ^2 \left(1+\eta ^2 k_1^2\right) }{k_1 k_3^3}\ ,
\eea
where in this specific case we have taken the $\tilde\d/\tilde\d\log k$ to represent $\d/\d\log k+3$, as if we were not to account for the non-rescaling of the spatial derivatives.
The reason for this mismatch is that in this naive consistency condition we are rescaling all the momenta, including the ones representing the external derivatives, that are not rescaled by the computation. By taking into account of this fact, the properly defined consistency condition holds:
\bea\label{eq:consistency_inside_derivative_correct}
&&\left\langle\frac{1}{a(\eta)^2}\left(\d_i\zeta\right)_{k_1}(\eta)\left(\d_i\zeta\right)_{k_2}(\eta)\zeta_{k_3}(\eta)\right\rangle\simeq- (2\pi)^3\delta^{(3)}(\vec k_1+\vec k_2+\vec k_3)\\\nonumber
&&\frac{1}{a(\eta)^2}\frac{\tilde\d\left[k_1^2P(k_1,\eta)\right]}{\tilde\d\log k_1}P(k_3)=-\frac{ H^6}{8 \mpl^4 \epsilon ^2}\cdot\frac{\eta ^4\,k_1}{k_3^3}\ , \qquad\quad\ k_1\ll k_3\ , \ \nonumber 
\eea
where now $\tilde\d/\tilde\d\log k$ is defined as in (\ref{eq:dtilde}) with the dimensions in length of the operator.

\end{appendix}

 \begingroup\raggedright\endgroup


\begin{thebibliography}{10}

\bibitem{Maldacena:2002vr}
  J.~M.~Maldacena,
  ``Non-Gaussian features of primordial fluctuations in single field
  inflationary models,''
  JHEP {\bf 0305} (2003) 013
  [arXiv:astro-ph/0210603].


\bibitem{Creminelli:2004yq}
  P.~Creminelli and M.~Zaldarriaga,
  ``Single field consistency relation for the 3-point function,''
  JCAP {\bf 0410} (2004) 006
  [arXiv:astro-ph/0407059].

\bibitem{Cheung:2007sv}
  C.~Cheung, A.~L.~Fitzpatrick, J.~Kaplan and L.~Senatore,
  ``On the consistency relation of the 3-point function in single field
  inflation,''
  JCAP {\bf 0802} (2008) 021
  [arXiv:0709.0295 [hep-th]].


\bibitem{Cheung:2007st}
  C.~Cheung, P.~Creminelli, A.~L.~Fitzpatrick, J.~Kaplan and L.~Senatore,
  ``The Effective Field Theory of Inflation,''
  JHEP {\bf 0803} (2008) 014
  [arXiv:0709.0293 [hep-th]].\\
  L.~Senatore and M.~Zaldarriaga,
  ``The Effective Field Theory of Multifield Inflation,''
  arXiv:1009.2093 [hep-th], to appear in JHEP.



\bibitem{Senatore:2009cf}
  L.~Senatore and M.~Zaldarriaga,
  ``On Loops in Inflation,''
  JHEP {\bf 1012} (2010) 008
  [arXiv:0912.2734 [hep-th]].


 \bibitem{eft}
  L.~Senatore, K.~M.~Smith and M.~Zaldarriaga,
   ``Non-Gaussianities in Single Field Inflation and their Optimal Limits from
  the WMAP 5-year Data,''
  JCAP {\bf 1001}, 028 (2010)
  [arXiv:0905.3746 [astro-ph.CO]];\\
  L.~Senatore, M.~Zaldarriaga,
  ``A Naturally Large Four-Point Function in Single Field Inflation,''
  JCAP {\bf 1101 } (2011)  003.
  [arXiv:1004.1201 [hep-th]];\\
  P.~Creminelli, M.~A.~Luty, A.~Nicolis and L.~Senatore,
  ``Starting the universe: Stable violation of the null energy condition and
  non-standard cosmologies,''
  JHEP {\bf 0612} (2006) 080
  [arXiv:hep-th/0606090];\\
  N.~Bartolo, M.~Fasiello, S.~Matarrese, A.~Riotto,
  ``Large non-Gaussianities in the Effective Field Theory Approach to Single-Field Inflation: the Bispectrum,''
  JCAP {\bf 1008 } (2010)  008.
  [arXiv:1004.0893 [astro-ph.CO]];\\
  N.~Bartolo, M.~Fasiello, S.~Matarrese, A.~Riotto,
  ``Large non-Gaussianities in the Effective Field Theory Approach to Single-Field Inflation: the Trispectrum,''
  JCAP {\bf 1009 } (2010)  035.
  [arXiv:1006.5411 [astro-ph.CO]];\\
  P.~Creminelli, G.~D'Amico, M.~Musso, J.~Norena, E.~Trincherini,
  ``Galilean symmetry in the effective theory of inflation: new shapes of non-Gaussianity,''
  JCAP {\bf 1102 } (2011)  006.
  [arXiv:1011.3004 [hep-th]];\\
  D.~Baumann, L.~Senatore, M.~Zaldarriaga,
  ``Scale-Invariance and the Strong Coupling Problem,''
  JCAP {\bf 1105}, 004 (2011).
  [arXiv:1101.3320 [hep-th]];\\
  D.~Baumann, D.~Green,
  ``Equilateral Non-Gaussianity and New Physics on the horizon,''
  JCAP {\bf 1109 } (2011)  014.
  [arXiv:1102.5343 [hep-th]];\\
  D.~Baumann, D.~Green,
  ``Signatures of Supersymmetry from the Early Universe,''
  [arXiv:1109.0292 [hep-th]];\\
  D.~L.~Nacir, R.~A.~Porto, L.~Senatore, M.~Zaldarriaga,
  ``Dissipative effects in the Effective Field Theory of Inflation,''
  [arXiv:1109.4192 [hep-th]];\\
  S.~R.~Behbahani, A.~Dymarsky, M.~Mirbabayi and L.~Senatore,
  ``(Small) Resonant non-Gaussianities: Signatures of a Discrete Shift Symmetry
  in the Effective Field Theory of Inflation,''
  arXiv:1111.3373 [hep-th];\\
  D.~Baumann and D.~Green,
  ``A Field Range Bound for General Single-Field Inflation,''
  arXiv:1111.3040 [hep-th];



\bibitem{Dalal:2007cu}
  N.~Dalal, O.~Dore, D.~Huterer and A.~Shirokov,
  ``The imprints of primordial non-gaussianities on large-scale structure:
  scale dependent bias and abundance of virialized objects,''
  Phys.\ Rev.\  D {\bf 77} (2008) 123514
  [arXiv:0710.4560 [astro-ph]].


\bibitem{loops1}
 G.~Pimentel, L.~Senatore, M.~Zaldarriaga
  ``On Loops in Inflation III: time-independence of $\zeta$ correlators'', to appear.
 
 
\bibitem{Weinberg:2005vy}
  S.~Weinberg,
  ``Quantum contributions to cosmological correlations,''
  Phys.\ Rev.\  D {\bf 72} (2005) 043514
  [arXiv:hep-th/0506236].

\bibitem{Weinberg:2006ac}
  S.~Weinberg,
  ``Quantum contributions to cosmological correlations. II: Can these
  corrections become large?,''
  Phys.\ Rev.\  D {\bf 74} (2006) 023508
  [arXiv:hep-th/0605244].

      \bibitem{Huang:2006eha}
        X.~Chen, M.~x.~Huang and G.~Shiu,
        ``The inflationary trispectrum for models with large non-Gaussianities,''
        Phys.\ Rev.\  D {\bf 74} (2006) 121301
        [arXiv:hep-th/0610235].


\bibitem{Kahya:2010xh}
  E.~O.~Kahya, V.~K.~Onemli and R.~P.~Woodard,
  ``The Zeta-Zeta Correlator Is Time Dependent,''
  Phys.\ Lett.\  B {\bf 694} (2010) 101
  [arXiv:1006.3999 [astro-ph.CO]].

\bibitem{Seery:2008ax}
  D.~Seery, M.~S.~Sloth and F.~Vernizzi,
  ``Inflationary trispectrum from graviton exchange,''
  JCAP {\bf 0903} (2009) 018
  [arXiv:0811.3934 [astro-ph]].

\bibitem{Leblond:2010yq}
  L.~Leblond and E.~Pajer,
  ``Resonant Trispectrum and a Dozen More Primordial N-point functions,''
  JCAP {\bf 1101} (2011) 035
  [arXiv:1010.4565 [hep-th]].


     \bibitem{Ganc:2010ff}
      J.~Ganc and E.~Komatsu,
    ``A new method for calculating the primordial bispectrum in the squeezed
      limit,''
      JCAP {\bf 1012} (2010) 009
        [arXiv:1006.5457 [astro-ph.CO]].


\bibitem{Creminelli:2011rh}
  P.~Creminelli, G.~D'Amico, M.~Musso and J.~Norena,
  ``The (not so) squeezed limit of the primordial 3-point function,''
  JCAP {\bf 1111} (2011) 038
  [arXiv:1106.1462 [astro-ph.CO]].
  
  
\bibitem{Creminelli:2012ed}
  P.~Creminelli, J.~Norena and M.~Simonovic,
  ``Conformal consistency relations for single-field inflation,''
  arXiv:1203.4595 [hep-th].


\bibitem{Giddings:2010nc}
  S.~B.~Giddings and M.~S.~Sloth,
  ``Semiclassical relations and IR effects in de Sitter and slow-roll
  space-times,''
  JCAP {\bf 1101} (2011) 023
  [arXiv:1005.1056 [hep-th]].


\bibitem{Gerstenlauer:2011ti}
  M.~Gerstenlauer, A.~Hebecker and G.~Tasinato,
  ``Inflationary Correlation Functions without Infrared Divergences,''
  JCAP {\bf 1106} (2011) 021
  [arXiv:1102.0560 [astro-ph.CO]].



\bibitem{Senatore:2012nq}
  L.~Senatore and M.~Zaldarriaga,
  ``On Loops in Inflation II: IR Effects in Single Clock Inflation,''
  arXiv:1203.6354 [hep-th].












\end{thebibliography}
\end{document}